\documentclass[conference]{IEEEtran}
\IEEEoverridecommandlockouts
\usepackage{cite}
\usepackage[ruled]{algorithm2e}
\usepackage{amsmath,amssymb,amsfonts}
\usepackage{algorithmic}
\usepackage{array}
\usepackage{balance}
\usepackage{cellspace}
\usepackage{graphicx}
\usepackage{hhline}
\usepackage{makecell}
\usepackage{multirow}
\usepackage{textcomp}
\usepackage[table]{xcolor}
\usepackage{url}
\usepackage{subfigure}
\def\BibTeX{{\rm B\kern-.05em{\sc i\kern-.025em b}\kern-.08em
    T\kern-.1667em\lower.7ex\hbox{E}\kern-.125emX}}

\SetKwInput{KwFeature}{Input}
\SetKwInput{KwOutput}{Output}

\title{Demystify Adult Learning: A Social Network and Large Language Model Assisted Approach}

\makeatletter
\newcommand{\linebreakand}{%
  \end{@IEEEauthorhalign}
  \hfill\mbox{}\par
  \mbox{}\hfill\begin{@IEEEauthorhalign}
}
\makeatother

\author{
\IEEEauthorblockN{Fang Liu}
\IEEEauthorblockA{\textit{Singapore University of Social Sciences}\\
Singapore, 599494\\
liufang@suss.edu.sg}
\and
\IEEEauthorblockN{Bosheng Ding}
\IEEEauthorblockA{
\textit{Nanyang Technological University}\\
Singapore, 639798\\
bosheng001@e.ntu.edu.sg}
\and
\IEEEauthorblockN{Chong Guan$^*$}
\IEEEauthorblockA{\textit{Singapore University of Social Sciences}\\
Singapore, 599494\\
guanchong@suss.edu.sg}
\linebreakand
\IEEEauthorblockN{Zhang Wei}
\IEEEauthorblockA{\textit{Singapore Institute of Technology}\\
Singapore, 138683\\
wei.zhang@singaporetech.edu.sg}
\and
\IEEEauthorblockN{Dusit Niyato}
\IEEEauthorblockA{
\textit{Nanyang Technological University}\\
Singapore, 639798\\
dniyato@ntu.edu.sg}
\and
\IEEEauthorblockN{Justina Tan}
\IEEEauthorblockA{
\textit{Singapore University of Social Sciences}\\
Singapore, 599494\\
justinatanpi@suss.edu.sg}

\thanks{This work was supported in part by A*STAR under its MTC Programmatic (Award M23L9b0052), MTC Individual Research Grants (IRG) (Award M23M6c0113), the Ministry of Education, Singapore, under the Academic Research Tier 1 Grant (Grant ID: GMS 693), and SIT’s Ignition Grant (STEM) (Grant ID: IG (S) 2/2023 – 792).}
\thanks{$^*$ Corresponding author. Email: guanchong@suss.edu.sg.}
}

\begin{document}
\bstctlcite{IEEEexample:BSTcontrol}

\maketitle

\begin{abstract}
Adult learning is increasingly recognized as a crucial way for personal development and societal progress. It however is challenging, and adult learners face unique challenges such as balancing education with other life responsibilities. Collecting feedback from adult learners is effective in understanding their concerns and improving learning experiences, and social networks provide a rich source of real-time sentiment data from adult learners. Machine learning technologies especially large language models (LLMs) perform well in automating sentiment analysis. However, none of such models is specialized for adult learning with accurate sentiment understanding. In this paper, we present \texttt{A-Learn}, which enhances \underline{a}dult \underline{learn}ing sentiment analysis by customizing existing general-purpose LLMs with domain-specific datasets for adult learning. We collect adult learners' comments from social networks and label the sentiment of each comment with an existing LLM to form labelled datasets tailored for adult learning. The datasets are used to customize \texttt{A-Learn} from several base LLMs. We conducted experimental studies and the results reveal \texttt{A-Learn}'s competitive sentiment analysis performance, achieving up to 91.3\% accuracy with 20\% improvement over the base LLM. \texttt{A-Learn} is also employed for word cloud analysis to identify key concerns of adult learners. The research outcome of this study highlights the importance of applying machine learning with educational expertise for teaching improvement and educational innovations that benefit adult learning and adult learners.
\end{abstract}

\begin{IEEEkeywords}
social network, large language model, generative artificial intelligence, sentiment analysis, applied artificial intelligence, adult learning
\end{IEEEkeywords}

\section{Introduction}
Adult learning is essential for personal growth and societal advancement \cite{tennant2007psychology}. It involves the continuous acquisition of knowledge and skill throughout one's adulthood \cite{mezirow1991transformative}. This process contributes to a more competent workforce and better employment prospects, and positively affects the economy with better job opportunities and increased productivity. It also promotes social inclusion by supporting marginalized groups to participate more fully in society \cite{desjardins2020political}. Unlike traditional education, which is typically confined to formal learning in childhood and adolescence, adult learning brings unique challenges. Adult learners must balance their educational pursuits with other adult responsibilities, which interfere with their regular attendance of classes, timely completion of assignments, and overall focus on studies. Adult learners may also struggle with adapting to new learning environments and technologies, and face financial constraints \cite{guan2020artificial}. 

Therefore, it is vital to gather feedback from adult learners and understand their unique needs to enhance the adult learning experience \cite{shaik2022review}. Traditionally, educators use surveys, in-class discussions, and course evaluation forms to collect feedback, but these methods may yield limited and potentially biased responses due to low participation rates. To address this, researchers are increasingly turning to social networks where a large volume of real-time comments from the learners are available. Analyzing such data allows for examining the social impacts of adult learning and offering insights into learners' real-time opinions, experiences, and sentiments \cite{philp2022brokering}. 

However, the vast and ever-increasing volume of data from social networks makes traditional analysis, often manual, impractical. To overcome this challenge, researchers have studied machine learning (ML) models to automate data processing and analysis. Within ML, natural language processing (NLP) is particularly useful for analyzing social network data, where text is the dominant data modality. An important application of NLP is sentiment analysis, which accesses the emotional tone behind comments to infer sentiments such as \texttt{Positive}, \texttt{Neutral}, or \texttt{Negative}. NLP excels in such tasks, and recent advancements have led to the development of many large language models (LLMs) that are trained on huge text datasets for enhanced performance in sentiment analysis. Some LLMs are specifically tailored for certain tasks and platforms, e.g., Twitter-RoBERTa \cite{sirisha2022aspect}, trained on 58 million tweets, is highly effective for various tasks including sentiment analysis for Twitter comments.

Unfortunately, even advanced LLMs may not perform optimally across all subjects due to the broad nature of training data which may not cover specialized areas comprehensively. While an LLM might demonstrate high accuracy in specific applications, there is a risk of over-fitting, where the model cannot generalize well to new data. This leads to a significant increase in error rates when the model is applied to unfamiliar data. Additionally, the constant evolution of language, including the emergence of new buzzwords and acronyms commonly used by students on social media, presents a continuous challenge for pre-trained LLMs which may struggle to adapt to these changes.

In this paper, we propose a customized LLM tailored to adult learning which is crucial for effective analysis in this field. Utilizing a specialized dataset for adult learning is essential for providing domain-specific knowledge to the models and refining the performance on the targeted topics. Accurate labelling of this specialized dataset is important for extracting meaningful insights. However, sourcing such accurately labeled datasets presents a significant challenge for adult learning research. To address this, we introduce a system called \texttt{A-Learn} for \underline{a}dult \underline{learn}ing sentiment analysis. \texttt{A-Learn} utilizes social network comments enhanced with LLM and accordingly customizes LLM models for adult learning. We outline our main contributions as follows.
\begin{itemize}
    \item We collected an adult learning dataset from multiple social network platforms with over 55,000 comments from current and prospective adult learners;
    \item We generated two additional datasets: a large-scale dataset labelled by LLM for model fine-tuning and customization, and a small-scale dataset labelled manually by human experts for performance evaluation;
    \item We developed \texttt{A-Learn} to incorporate adult learning knowledge into existing LLMs with optimized training and produce LLMs customized for adult learning;
    \item We conducted experimental studies and demonstrated \texttt{A-Learn}'s competitive performance in sentiment analysis for adult learning, with up to 91.3\% accuracy and a 20\% improvement over the baseline.
\end{itemize}

The rest of the paper is organized as follows. The system architecture of \texttt{A-Learn} is presented in Section \ref{sec:system}. Section \ref{sec:method} describes our methodology in detail and Section \ref{sec:exp} presents an experimental study of the proposed solution as well as result discussions. Finally, we conclude the paper in Section \ref{sec:conclusion}.

\section{System Architecture} 
\label{sec:system}
\begin{figure}
  \centering  \includegraphics[width=0.48\textwidth]{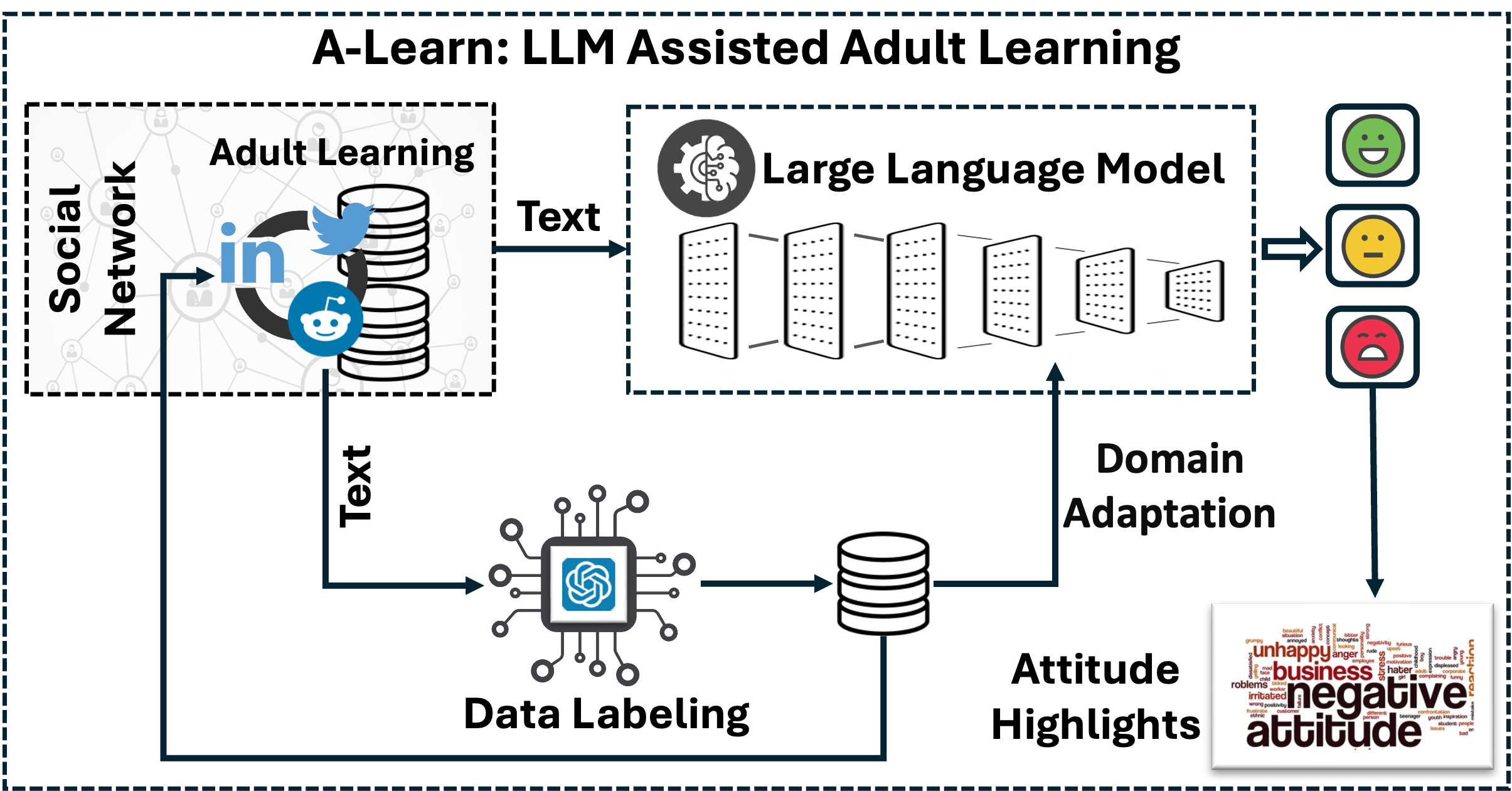}
  \caption{The system architecture of \texttt{A-Learn} for adult learning sentiment analysis. \texttt{A-Learn} is ML-based and processes the social network comments as input to generate sentiment labels as output. \texttt{A-Learn}'s sentiment analysis plays a crucial role in identifying key concerns and facilitating educational innovations in adult learning.
  }
\label{fig:sys-arch}
\end{figure}

In this section, we present the system architecture of \texttt{A-Learn} and an illustration of it is available in Fig. \ref{fig:sys-arch}.

\subsection{LLM for Sentiment Analysis}
The sentiment analysis of adult learning can be conducted in different ways and LLM is one of the effective solutions to automatic the analysis. In a typical scenario, an LLM takes student comments (e.g., from social media and questionnaires) as input. The data is pre-processed such as modality alignment before the ML analysis, where sentiment analysis is treated as a classification problem and the model's role is to generate sentiment labels including \texttt{Positive} for satisfaction, \texttt{Negative} for dissatisfaction, and \texttt{Neutral} otherwise. The sentiment labels allow for the grouping of comments, where in-depth investigation can be performed for each group to gain a deep understanding of adult learners for teaching quality enhancement and educational innovation.

\subsection{LLM Customization for Adult Learning}
Despite being a promising technology, existing LLM is not specialized for adult learning. There are significant differences between general-purpose text analysis and adult learning sentiments, where some phrases, acronyms, etc., are not common in generic settings. This calls for a customization, or domain adaptation, of existing LLM to adult learning scenarios. In \texttt{A-Learn}, we first collect an adult learning dataset from various social media platforms to capture the adult learning concerns in different aspects. Yet, no one has indicated the sentiment of any comments. Thus, we apply an existing LLM, ChatGPT in specific, to label the comments in the dataset. The labels may not be fully correct from the views of human experts, but they are cheap to obtain and very likely to be correct. Now, with the dataset specialized for adult learning as well as associated sentiment labels, we customize the generic LLMs to adapt to adult learning with accurate sentiment analysis. This is a typical transfer learning concept, which has been verified in other applications \cite{liu2023transline}. We expect the transfer or customization, to be effective with our specialized dataset and training strategies. 

\section{Methodology} 
\label{sec:method}
We present the detailed methodology of \texttt{A-Learn} in this section, for both data collection and knowledge adaptation.

\begin{figure}
	\centering
	\def \heightGran{1.25in}
	\subfigure[Mixed Sentiment]{\includegraphics[width=0.48\textwidth]{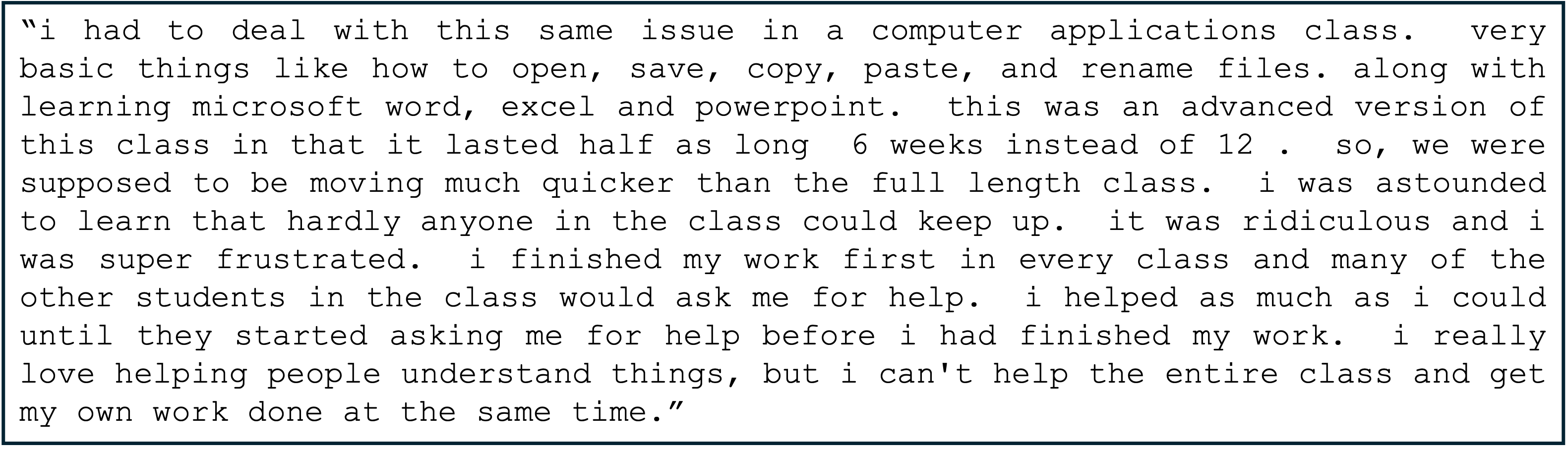}\label{figComMix}}\hspace{0em}
	\subfigure[Negative Sentiment]{\includegraphics[width=0.48\textwidth]{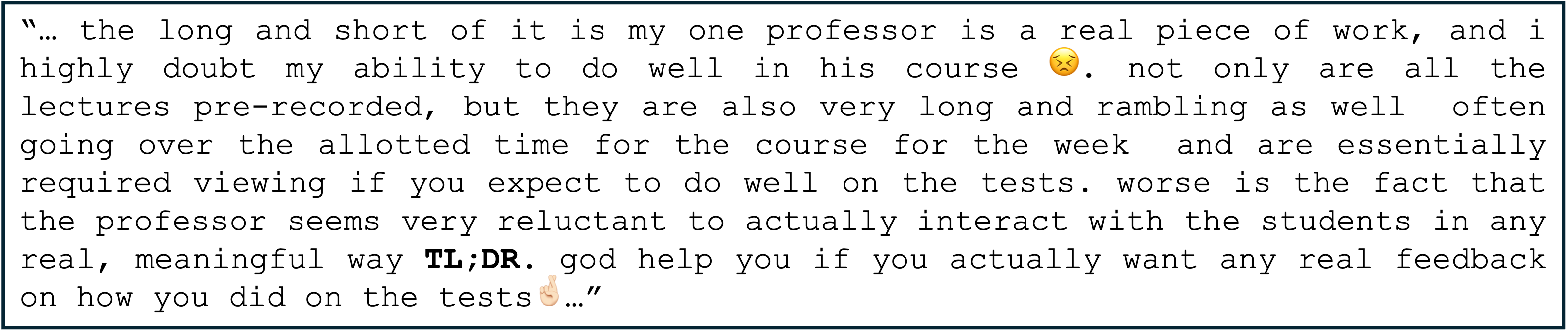}\label{figComNeg}}
	\caption{Sample comments with different emotions/sentiments about time management and teaching quality for adult learning.} 
	\label{fig:comment}
\end{figure}

\subsection{LLM-assisted Sentiment Label Acquisition}
Adult learning as a specific topic of sentiment analysis involves unique patterns that are not common in generic settings. Frequent phrases such as time management, work-life balance, teaching quality, and resource allocation, may not be well modeled in general-purpose LLMs. Besides, the sentiment can be complicated with mixed emotions. We show one sample comment in Fig. \ref{figComMix}, where a student expresses enthusiasm for helping peers, but feels struggles when additional demands arise and affect his/her own learning progress. Fig. \ref{figComNeg} presents a sample comment with negative sentiment, featuring network slang and emojis, which add analysis complexity.

To understand the unique patterns of adult learning, we consider social networks that offer comprehensive information regarding student comments about adult learning. However as mentioned above, we lack sentiment labels of the comments and face challenges to interpret the data and extract insights. As such, we introduce a labelling process, which shall not be fully manual considering the labelling complexity and cost. We follow a hybrid approach with the assistance of both LLM and human experts. We task ChatGPT to label comments with sentiment and let human experts verify the labels, assuming verification is much easier and faster than labelling. According to recent studies \cite{ding2023gpt,li2023coannotating} for other applications, such an approach can be cost-efficient with big time-savings. 

\subsection{\texttt{A-Learn} with LLM Customization} 
\begin{figure}
  \centering  \includegraphics[width=0.48\textwidth]{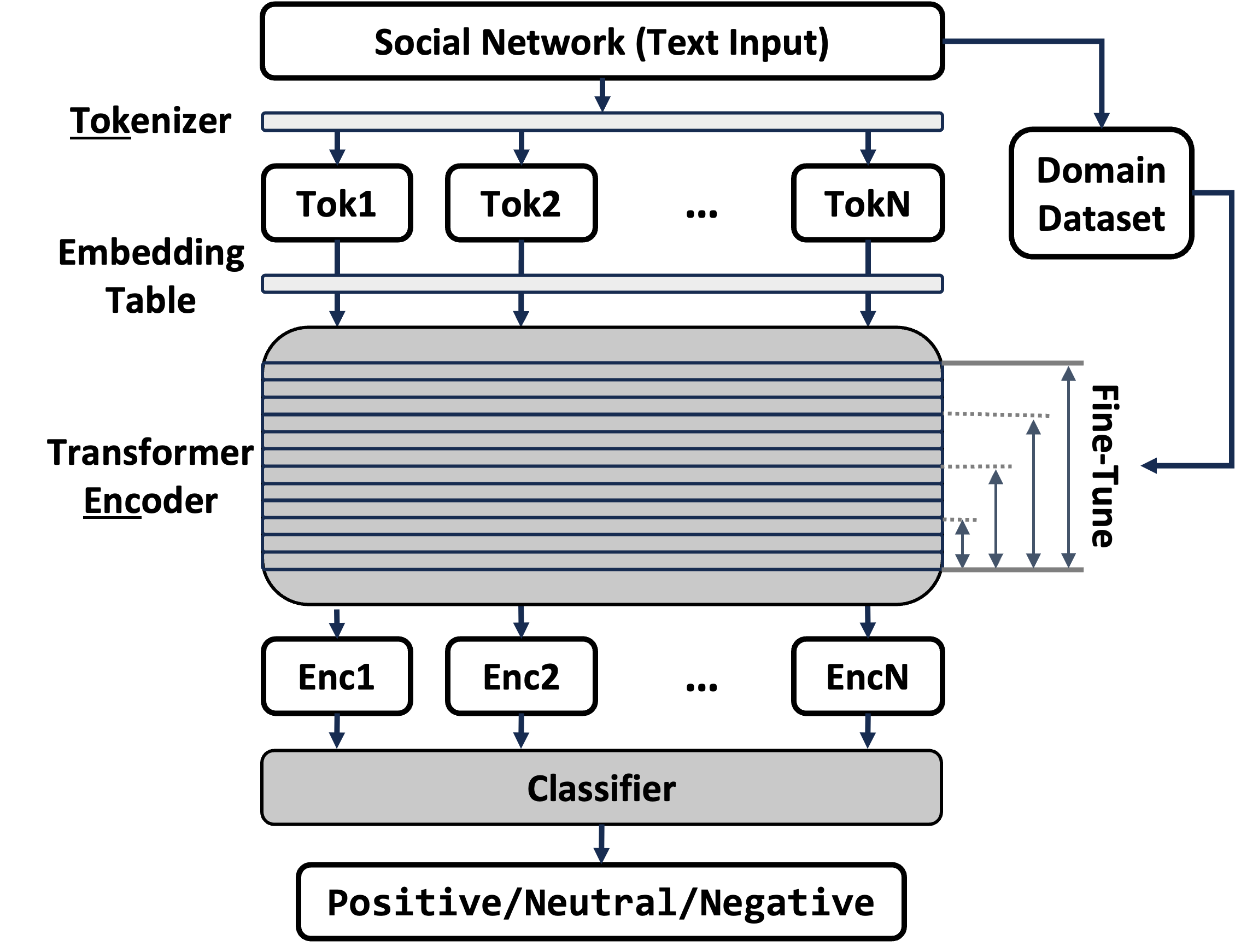}
  \caption{An illustration of \texttt{A-Learn} with social network comments as the input. Domain knowledge is extracted from the comments and represented as sentiment initiates, used to fine-tune different layers in LLM's transformer encoder. The customized model can identify the sentiment of a comment.
  }
\label{fig:model}
\end{figure}

It takes a huge dataset to train an LLM to achieve optimal performance, but a huge dataset is often very costly to obtain. This is no difference for adult learning and we shall choose the data-lite strategies. We first adopt pre-trained LLMs with optimal performance for understanding generic text content. Each LLM consists of many parameters and the optimal values for generic usage are known and downloadable for these LLMs. We take such optimal values as the starting point for customizing the parameters to adult learning. We adjust and optimize the values as a model re-training or fine-tuning process, and our ultimate goal is to adopt the parameter values for general-purpose text analysis to the domain of adult learning. This process is illustrated in Fig. \ref{fig:model}. LLM tokenizes the input text and converts them into numerical representations, which are then fed into the transformer encoder for language comprehension. We follow transfer learning to re-train or fine-tune the models with the adult learning dataset. This process adjusts the weights of various layers in the transformer architecture. Specifically, we can freeze some layers near the input and fine-tune the rest layers near the output. The more layers we freeze, the more information from the pre-trained model is retained. Conversely, adjusting more layers allows the penetration of more domain-specific information, which however demands more data to perform well. The output from the transformer encoding is passed to a classifier, which produces the final three-fold sentiment labels, \texttt{Positive}, \texttt{Negative}, and \texttt{Neutral}. 

Let $\mathcal{M}(x;\theta)$ be the sentiment analysis modal which analyzes the comment $x$ and determines its sentiment label as $y=\mathcal{M}(x;\theta)$ with model parameters $\theta$. We assume two connected stages of the whole model as $(\mathcal{M}_\text{f},\mathcal{M}_\text{t})$ where $\mathcal{M}_\text{f}$ and $\mathcal{M}_\text{t}$ are the sets of frozen and trainable layers, respectively, along with the corresponding model parameters $(\theta_{\text{f}},\theta_\text{t})$. In \texttt{A-Learn}, we initialize the parameters as $\theta_\text{f}^0$ and $\theta_\text{t}^0$ which are copied from the optimal parameter values from the base LLMs. The aim of customization is to find the optimal parameter values $\theta_\text{t}^*$ for adult learning sentiment analysis as,
\begin{equation}
\theta_\text{t}^* = \arg\min_{\theta_\text{t}} \mathcal{L}(\mathcal{D}; \theta_\text{f}^0,\theta_\text{t}),
\end{equation}
where $\mathcal{L}(\cdot)$ is the cross-entropy loss function, $\mathcal{D}$ is our labelled dataset, and $\theta_\text{f}$ remains unchanged as $\theta_\text{f}^0$. The final model of \texttt{A-Learn} is $\mathcal{M}(\cdot;\theta^*)$ where $\theta^*=(\theta_\text{f}^0, \theta_\text{t}^*)$. Specifically, re-training means all transformer layers are trainable where $\mathcal{M}_\text{f}$ is null and the optimal model is $\mathcal{M}(\cdot;\theta^*)$ where $\theta^*= \theta_\text{t}^*$. With comments and \texttt{A-Learn}'s generated labels, educators can perform in-depth analysis for improved teaching.

\section{Experimental Study} 
\label{sec:exp}
We present our experimental study in this section to demonstrate \texttt{A-Learn}'s sentiment analysis performance for adult learning and interpret the experiment results to find insights. \texttt{A-Learn} is data-driven and we start with data description. 

\subsection{Dataset and Data Pre-processing}
\label{subsec:data}
In this part, we first introduce the data collection for our experiments, followed by data pre-processing descriptions. 

\subsubsection{Data Collection}
\label{subsubsec:source}
Social networks are popular among various groups of users. It provides rich and extensive data that facilitates in-depth analysis of human opinions and sentiments. In \texttt{A-Learn}, we scrap plenty of comments related to adult learning from multiple social media platforms, including Reddit, Twitter, and Tumblr. They are selected because of their diverse user demographics (e.g. various age groups and backgrounds) and well-supported APIs. Our comment search is based on keywords such as \texttt{adult learning}, \texttt{part-time studies}, \texttt{continuous learning}, and \texttt{work study}. In total, we have 31,724 Reddit comments collected from 2017 to 2023, 22,270 Twitter tweets in 2023, and 1,043 Tumblr comments from 2011 to 2023.

\subsubsection{Data Pre-processing}
The raw data (comments) collected from the above stage is not ready for ML models and necessary pre-processing is needed to facilitate the ML-based analysis. Firstly, we remove duplicated and incomplete content. Then, we scan the content and eliminate special text components, including HTML links and tags, emojis, and newline characters. All contents are normalized and converted to lowercase to ensure consistency. Finally, the contents are processed through tokenization, lemmatization, and stop-word removal to prepare them to be compatible with ML.

\subsubsection{Datasets for Experiments}
We present two datasets specialized for adult learning with different labelling schemes. 

\paragraph{LLM Labels for Training} 
We use LLM to generate a relatively large-scale labelled dataset. Particularly, we use \texttt{GPT-3.5} and get its opinion about the sentiment of each of the 1,000 randomly selected social network comments. Some labels are randomly selected for manual inspection by human experts, to evaluate if the labelling quality meets our expectations. Note that we do not expect fully correct labels in \texttt{A-Learn} training. 

\paragraph{Expert Labels for Testing} 
We need fully correct labels for performance evaluation. Thus we introduce human expert support. We randomly select comments from our collected dataset. We let one human expert read each comment and assign a sentiment label to it, and the label is verified by two other human experts. Only if the three experts reach a consensus, we include the comment along with its label into the testing dataset. We stop the labelling once we have 300 labelled comments with one-third for each category of sentiment. This will be our testing dataset.

\subsection{\texttt{A-Learn} Performance Analysis} 
\label{subsec:perfromance}
\texttt{A-Learn} considers several widely used or latest LLMs as base models and then customizes them with our labelled adult learning dataset. We first present the experimental setup. 

\subsubsection{Experimental Setup}
We use Google Colab to implement \texttt{A-Learn}. For the base model, we select the widely used BERT \cite{kenton2019bert} and five of its variants, including ALBERT \cite{lan2019albert}, RoBERTa \cite{liu2019roberta}, DistilBERT, DistilBERT-SST2 \cite{sanh2019distilbert}, and Twitter-RoBERTa \cite{sirisha2022aspect}. Here, DistilBERT-SST2 and Twitter-RoBERTa are two specialized models trained with the SST2 dataset (a benchmark for sentiment analysis) and a Twitter dataset, respectively. Some models have different versions and we choose the default one, normally with a suffix of \texttt{-base}, which is excluded in this paper for simplicity. 

In the following part, we use \texttt{Base} as a reference of the original LLMs and \texttt{A-Learn} is the LLM customized for adult learning. \texttt{A-Learn} is fine-tuned with the LLM-labelled dataset for at least ten epochs. Accuracy, in the percentage of the number of correct predictions to the total number of predictions, is used as the main performance metric based on the testing results on expert-labelled dataset. All experimental procedures are executed on NVIDIA V100 GPUs.

\begin{table}
\renewcommand{\arraystretch}{1.3}
\centering
\caption{The performances in accuracy (\%) and training time in minute of \texttt{Base} models and \texttt{A-Learn} customized from the models. \texttt{A-Learn} always achieves better accuracy compared to the corresponding \texttt{Base} models with significant improvement and manageable time usage for training.}
\begin{tabular}{@{\extracolsep{6pt}}ccccc@{}}
\hline\hline
Model Name & \texttt{Base} & \texttt{A-Learn} & Improvment & Time\\\hline
Bert & 41.0 & 61.7 & 20.7 & 11 \\ \hline
ALBERT & 33.0 & 35.7 & 2.7  & 10 \\ \hline
RoBERTa & 32.0 & 56.3 & 24.3 & 12 \\ \hline
Distilbert & 36.0 & 59.0 & 23.0 & 6  \\ \hline
Distilbert-SST2 & 44.0 & 59.3 & 15.3 & 7 \\ \hline
Twitter-RoBERTa & 76.3 & \textbf{91.3} & 15.0 & 12\\ \hline
\hline
\end{tabular}
\label{tab:accuracy}
\end{table}

\subsubsection{Sentiment Analysis Performance}
We evaluate \texttt{A-Learn}'s performance on improving pre-trained \texttt{Base} models for adult learning sentiment analysis. The statistical results of our experiments are presented in Table \ref{tab:accuracy}. The results demonstrate a consistent and substantial accuracy improvement of \texttt{A-Learn} with six different models after model customization with domain-specific data. The accuracy improvement ranges from 2.7\% to 24.3\% and the average is 16.8\%. This validates our hypothesis that incorporating domain-specific knowledge into existing models can boost the model's performance in adult learning sentiment analysis. Notably, \texttt{A-Learn} based on Twitter-RoBERTa reaches an accuracy of 91.3\%, with a big increase from 76.3\% when Twitter-RoBERTa is not customized. This underscores the importance of customizing the base models with the data of a similar format (e.g., from Twitter) for sentiment analysis. 

\texttt{A-Learn} is also efficient. The training time is about ten minutes for all tested LLMs. Compared with training a general-purpose model which takes days or even weeks, \texttt{A-Learn} is much faster. One reason is that \texttt{A-Learn} starts its training with the optimal model parameters for generic applications that share similarities with adult learning. This leads to reduced training difficulty and accelerated training.

\subsection{Sensitivity Analysis}
In this section, we explore different \texttt{A-Learn} settings and identify the optimal configuration of \texttt{A-Learn}.

\subsubsection{Number of Trainable Layers} 
We have proved that customization is effective for adult learning sentiment analysis. In this part, we further investigate how to customize the knowledge to achieve optimal performance. We take the best-performing model, Twitter-RoBERTa, as the \texttt{Base} model and we conduct sensitivity analysis. We examine \texttt{A-Learn}'s performance sensitivity to the number of trainable layers, and we show the results in Table \ref{tab:layer}. This analysis aims to identify the ideal configuration of the trainable layers to reach the best performance of adult learning sentiment analysis.

Specifically, we investigate the transformer stage of Twitter-RoBERTa, which comprises 12 layers. Generally, the layers closer to the model output are more application-specific, and should be trainable to adapt to our application. In our experiment, let $n$ be the number of trainable layers nearest to the model output, while the remaining layers are frozen. For example, $n=0$ refers to the \texttt{Base} model where no layers is trainable. In \texttt{A-Learn}, we explore different settings of $n$ ranging from 1 to 12. Due to the limit of space, we report the representative ones with the minimum $n=1$ layer, about half of the layers with $n=7$, and all the layers with $n=12$.

The results show an increase in accuracy as the number of trainable layers $n$ increases, reaching a peak accuracy of 91.3\% when all layers are trainable. This trend suggests that limiting the number of trainable layers can negatively impact the model’s ability to adapt to a specific task, potentially due to insufficient adjustment to the unique patterns of our task. Notably, the accuracy drops to 72\% when $n=1$, even lower than the performance of \texttt{Base}. This implies insufficient domain adaptation for the new application, while the analysis of the original model is disrupted with model changes. In conclusion, allowing more layers to be trainable helps \texttt{A-Learn} better adapt to the sentiment analysis application of adult learning.

\begin{table}
\renewcommand{\arraystretch}{1.3}
\centering
\caption{Performance sensitivity of \texttt{A-Learn} to different numbers of trainable layers in terms of accuracy (\%). The improvement column refers to the \texttt{A-Learn}'s accuracy improvement in percentage over the base model.}
\begin{tabular}{@{\extracolsep{6pt}}cccc@{}}
\hline\hline
Model  & No. of Trainable Layers & Accuracy & Improvement \\\hline
\texttt{Base}   & 0 (0\%)   & 76.3    &  $-$  \\\hline
\multirow{3}{*}{\texttt{A-Learn}} & 1 (8\%)  & 72.0    & -5.6 \\\cline{2-4}
    & 7 (58\%)  & 83.7    & 9.7 \\\cline{2-4}
    & 12 (100\%) & \textbf{91.3}    & 19.7  \\
\hline\hline
\end{tabular}
\label{tab:layer}
\end{table}

\subsubsection{Different LLM-assisted Data} 
In sentiment analysis, there are two primary approaches using LLMs for data labelling: 1) assign sentiment labels to comments, and 2) generate comments given specific labels. Let a comment be $x$. The role of LLM in the first approach is to generate $\hat{y}=\text{LLM}(x)$ where $x$ is the input (real data) of the LLM and $\hat{y}$ is LLM's inferred sentiment label (not ground-truth). We represent this approach as $x \rightarrow \hat{y}$. For the second approach, represented by $y \rightarrow \hat{x}$, we provide the sentiment label $y$ (without adding any comments), and let the model generate a list of comments corresponding to the label, i.e., $\hat{x}=\text{LLM}(y)$ for each synthetic comment $\hat{x}$. We select Twitter-RoBERTa as the \texttt{Base} and report \texttt{A-Learn}'s performance in Table \ref{tab:label}. 

Our default setting of LLM-assisted labelling is shown in the last line of $y \rightarrow \hat{x}$, where \texttt{A-Learn} is 91.3\% accurate. Compared with the base model without customization, it is nearly 20\% more accurate. With another scheme $y \rightarrow \hat{x}$, the accuracy declines compared to the base model, by 15.6\% with 64.3\% sentiment analysis accuracy. A potential reason is that social network comments usually are complex and the texts are often unstructured and informal. There are also new and non-common phrases with the use of slang, abbreviations, and various languages or dialects. Generating synthetic comments $\hat{x}$ purely from labels $y$ with LLM is challenging. We notice that LLM tends to standardize the output comments with similar language styles, length, etc., where the complexity of the real comments is not well captured. As a result, customization with such data does not lead to improved performance for \texttt{A-Learn}, and even worse, the accuracy drops. We assume that the dataset fails to bring extra value in adult learning to \texttt{A-Learn} and meanwhile disrupts the base model with optimal performance for general-purpose applications.

In summary, our experiments demonstrate that the LLM-labelled dataset is useful for customizing the general-purpose LLM to adult learning with optimized configurations. This aligns with the findings from previous works \cite{ding2023gpt} that tagging unlabeled data using LLMs is an effective method compared to traditional data collection schemes. This is especially true for our application where the comments own unique features specialized to adult learning.

\begin{table}
\renewcommand{\arraystretch}{1.3}
\centering
\caption{The performances in accuracy (\%) of \texttt{Base} models and \texttt{A-Learn} for sentiment analysis with different labelling methods. The improvement in percentage is based on the comparison with \texttt{Base} model.}
\begin{tabular}{@{\extracolsep{6pt}}ccc@{}}
\hline\hline
Customization Data  & Accuracy & Improvement (\%) \\\hline
\texttt{Base}  & 76.3    &  -  \\\hline
LLM $y \rightarrow \hat{x}$ & 64.3 & -15.6 \\\hline
LLM $x \rightarrow \hat{y}$ (ours) & \textbf{91.3} & 19.7 \\\hline
\hline
\end{tabular}
\label{tab:label}
\end{table}

\subsection{\texttt{A-Learn}-enabled Insights Visualization} 
\begin{figure}
  \centering  \includegraphics[width=0.48\textwidth]{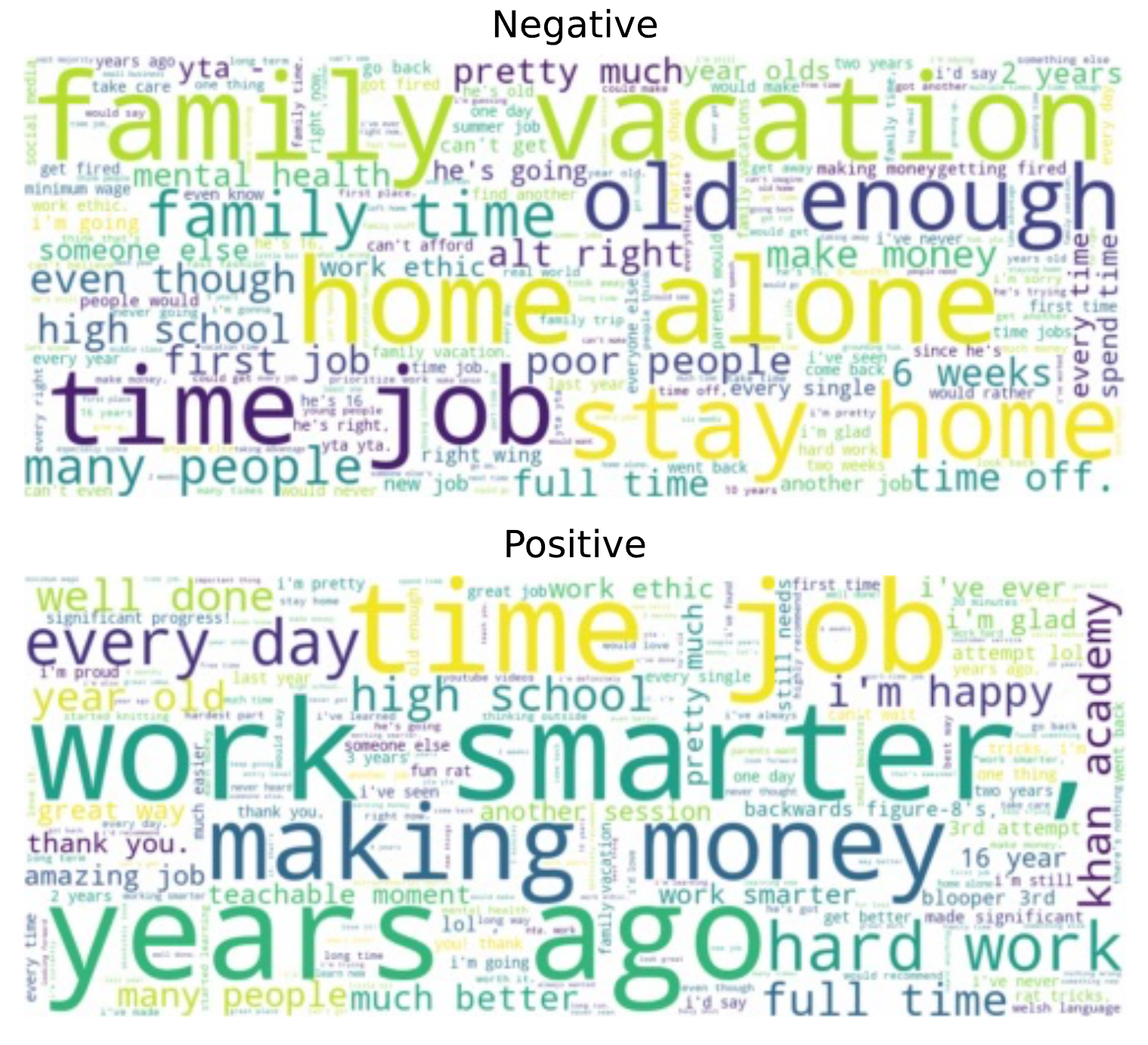}
  \caption{The word clouds with highlighted adult learning concerns for both negative and positive sentiments. Larger words appear more frequently in the comments and imply significant concerns of adult learners.}
\label{fig:WordCloud}
\end{figure}

In this part, we apply \texttt{A-Learn} in its optimal configuration for analyzing the comments collected from all three platforms, Reddit, Twitter, and Tumblr. We apply \texttt{A-Learn} on the collected comments that are not used in neither the training nor testing dataset of small-scale, meaning most of the comments collected are used in this part. After \texttt{A-Learn} processing, all these comments are assigned with a sentiment label. We pick out the comments with the same label, specifically \texttt{Positive} and \texttt{Negative}. For each group of comments with the same label, we conduct a word cloud analysis to visualize the most frequent words for each sentiment of adult learning. We show the word cloud figures in Fig. \ref{fig:WordCloud} where the words with limited linguistic information like stop words are excluded.

\subsubsection{Negative}
Negative sentiment is important as it often sparks insights into teaching improvement and educational innovation. As seen from the figure, negative comments are largely personal with domestic challenges. Unlike traditional learners, adult learners must balance their studies with family and work responsibilities. This is evidenced by the highlighted words such as \texttt{time job}, \texttt{family vacation}, \texttt{family time}, etc. Additionally, adult learners often lack frequent interactions with peers and instructors and spend much of their study time \texttt{home alone}. Older students may face difficulties related to their age and physical fitness, indicated by terms such as \texttt{old enough}, while younger students may experience pressure related to their \texttt{first job} and financial obligations such as \texttt{make money}.

\subsubsection{Positive}
Positive sentiment is related to professional development and advice for effective learning. For instance, adult learners emphasize the need for \texttt{work smarter} and \texttt{work hard} at their jobs, to free up time for adult learning. This highlights the continued importance of balancing work and study. Adult learning shall be persistent, with phrases \texttt{every day}. Interestingly, utilizing online resources and open-sourced tools such as \texttt{khan academy} seems to be effective in enhancing adult learning. Overall, \texttt{A-Learn} based word clouds for different sentiments can shed light on teaching improvement, policy adjustment, etc., with special attention on adult learning rather than traditional learning. 

\section{Conclusion} \label{sec:conclusion}
In this paper, we introduced \texttt{A-Learn} to demystify the sentiment of adult learners by analyzing their social network comments. \texttt{A-Learn} leveraged advanced LLMs and was tailored for adult learning. It can be based on different LLMs and several of them were investigated in this paper. \texttt{A-Learn} customizes the base LLMs, often general-purposed, with adult learning dataset extracted from social networks and labelled by an LLM. Experiment results demonstrated that customizing existing LLMs with domain-specific data is effective, and \texttt{A-Learn} can achieve high accuracy in sentiment analysis in adult learning scenarios. Specifically, \texttt{A-Learn} achieves its peak performance when Twitter-RoBERTa is used as its \texttt{Base} model. Its accuracy can be up to 91.3\%, surpassing the non-customized base model by 19.7\%. Besdies, customization itself demands an optimized process, and we identified the optimal configuration of \texttt{A-Learn} training through sensitivity studies. Overall, this is an interdisciplinary research combining the strengths of ML and social science. The research outcome holds substantial implications for continuing education and insights for better-supporting adult learners.


\section*{Acknowledgement}
The authors acknowledge the contribution of the undergraduate students from the Singapore Institute of Technology for preliminary study in integrative team projects in the AY22/23.

\bibliographystyle{IEEEtran}
\bibliography{reference}

\end{document}